\documentclass[a4paper]{article}

\usepackage{INTERSPEECH2021}
\usepackage{url}

\title{Information Sieve: Content Leakage Reduction in End-to-End Prosody Transfer for Expressive Speech Synthesis}
\name{Xudong Dai$^{1,\dagger}$, Cheng Gong$^{1,\dagger}$, Longbiao Wang$^{1,2*}$, Kaili Zhang$^{1}$}

\address{
$^1$ Tianjin Key Laboratory of Cognitive Computing and Application, \\
College of Intelligence and Computing, Tianjin University, Tianjin, China \\
$^2$ Tianjin University-Huiyan Technology Joint AI Lab, Tianjin University, Tianjin, China
}
\email{\{hsudongdai, gongchengcheng,  longbiao\_wang, kellyzhang\}@tju.edu.cn}

\begin{document}

\maketitle

\begin{abstract}
Expressive neural text-to-speech (TTS) systems incorporate a style encoder to learn a  latent embedding as the style information. However, this embedding process may encode redundant textual information. This phenomenon is called content leakage. Researchers have attempted to resolve this problem by adding an ASR or other auxiliary supervision loss functions. In this study, we propose an unsupervised method called the ‘‘information sieve’’ to reduce the effect of content leakage in prosody transfer. The rationale of this approach is that the style encoder can be forced to focus on style information rather than on textual information contained in the reference speech by a well-designed downsample--upsample filter, i.e., the extracted style embeddings can be downsampled at a certain interval and then upsampled by duplication. Furthermore, we used instance normalization in convolution layers to help the system learn a better latent style space. Objective metrics such as the significantly lower word error rate (WER) demonstrate the effectiveness of this model in mitigating content leakage . Listening tests indicate that the model retains its prosody transferability compared with the baseline models such as the original GST-Tacotron and ASR-guided Tacotron.\footnotetext{\ $^\dagger$These authors contributed to the work equally and should be regarded as co-first authors. * marks the corresponding author.}
\end{abstract}

\section{Introduction}
In recent years, several end-to-end neural text-to-speech models have significantly improved the quality and naturalness of synthesized speech \cite{dblp:conf/interspeech/WangTacotron, dblp:conf/icassp/ShenPWSJYCZWRSA18, dblp:conf/nips/RenRTQZZL19, dblp:conf/iclr/deepvoice3,  gong2021improving}. However, the style of speech synthesized by these models is strongly coupled with the training dataset, which typically converges to a neutral prosody, affecting the expressiveness and naturalness of synthesized speech.

Hence, neural prosody transfer (PT) techniques were introduced that added style information to the synthesized audio from a reference audio. Several approaches were proposed for this task. For example, Skerry-Ryan et al. \cite{dblp:conf/icml/towardse2e} conditioned the model with a style encoder to capture sentence-level time-independent features like emotion and style. This method is called the coarse-fined prosody transfer (CPT) model. \cite{dblp:conf/icml/styletokens, dblp:conf/icassp/ZhangPHL19, dblp:conf/interspeech/AkuzawaIM18} followed a similar methodology. Based on the CPT models, Hsu et al. \cite{dblp:conf/iclr/HsuHierarchical} proposed a fine-grained prosody transfer model to focus on low-level time-dependent features such as stress, rhythm, melody, etc. This methodology was followed by \cite{dblp:conf/interspeech/KlimkovHierarchical,dblp:conf/icassp/SunHierarchical, dblp:conf/interspeech/HonoHierarchical}. 
\begin{figure*}[ht]
  \centering
  \includegraphics[width=15cm]{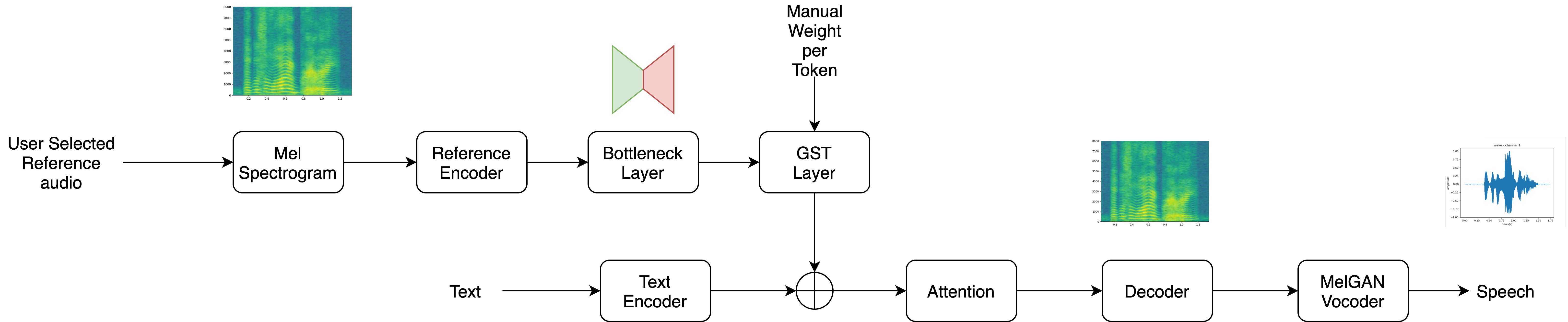}
  \caption{Model Structure for prosody transfer. We insert our sieve layer between the style encoder and GST layer. The output of GST layer is concatenated with the output of the text encoder. During inference, users could choose any reference audio or manually set weight per token.}
  \label{fig:modelstrcuture}
\end{figure*}
However, both methodologies extract a latent style representation from a mel-spectrogram without supervision. The mel-spectrogram, however, is both the input of the style encoder and the target of training. That is, the style encoder might encode some textual information if its capacity exceeds its mission requirement. This phenomenon is called content leakage\cite{dblp:conf/icassp/HuMutualInformation}, and it results in flaws in the synthesized audio, including blurred or missed words. Liu et al. \cite{dblp:conf/slt/asrguidedtacotron} used an ASR model to judge the training output, forcing the GST-Tacotron model \cite{dblp:conf/icml/styletokens} to learn to resolve unpaired text and reference audio with an additional ASR loss. Gururani et al. \cite{dblp:journals/corr/gururani} decomposed the reference speech into several low-dimensional features as the input to the style encoder instead of mel-spectrograms. Hu et al.\cite{dblp:conf/icassp/HuMutualInformation} attempted to minimize the mutual information between style embedding and text embedding to purify the former.

Content leakage is a challenge in many other research fields involving style transfer. Song et al. \cite{dblp:conf/ccs/SongR20} investigated this problem in several natural language processing (NLP) models, indicating the surplus ability of modern embedding models to express. Ridgeway et al. \cite{dblp:journals/corr/cvcontentleakage} used an additional \textit{Leakage Filtering Loss} to penalize a one-shot style transfer model for leaked content during training. Roy et al. \cite{dblp:conf/cvpr/RoyContentLeakage} used a new method called \textit{Maximum Entropy Adversarial Representation Learning} to minimize information leakage about a given sensitive attribute, exploring the potential of adversarial learning. Similarly, Hu et al. \cite{dblp:conf/icassp/HuMutualInformation} also proved the validity of minimizing mutual information. Generally, these methods aim to tackle this problem by adding supervision; however, the supervisory components are computationally taxing and possibly inaccurate.

We proposed a new technique called \textit{Information Sieve}, which can address content leakage while retaining the simplicity of the model and similarity of the transferred prosody with only a reconstruction loss, inspired by a voice conversion system called AutoVC\cite{dblp:conf/icml/QianAutoVC}. The proposed system’s workflow is the same as that of GST-Tacotron, except for a new attachment to its style encoder. Specifically, we used a downsample--upsample structure to filter unwanted textual information from the reference output to condition the vanilla Tacotron\cite{dblp:conf/interspeech/WangTacotron}. Speech recovered from generated mel-spectrogram sounds natural and is as expressive as the reference speech. Objective evaluations demonstrate the effective mitigation of content leakage. The main contributions of this study are summarized below.

\begin{enumerate}
\item We demonstrate the capacity of the style encoder is superfluous with similar quality of synthesized audio and lower WER of our models. A \textit{Sieve} to filter the redundant information without requiring an adequate output dimension for encoders to mitigate the content leakage.
\item We demonstrated that this method helps the GST layer assign more weights to certain tokens, compared with the approximately average distribution of weights by the original GST-Tacotron. This indicates this method helps each style token better capture its own style.
\item We used the modified style encoder’s output, concatenated with text embedding, as inputs for a Tacotron decoder, denoted as information sieve GST (IS-GST). Experiments demonstrated the effective mitigation of content leakage with the preservation of its naturalness and expressiveness during prosody transfer.
\end{enumerate}

\section{Sieve Prosody Transfer Architecture}
This approach leverages prior research on the encoding capacity of embedding models, indicating that the encoders can encode more information than the mission requirement, disturbing the decoder’s workflow. Specifically, the method filters the content encoder of an AutoVC, the inspiration for this approach, using an information sieve. Therefore, we hypothesize that style embedding could leak textual information due to its superfluous capacity. Further, the imposition of an adequate limit on the style encoder could help generate pure style embeddings, which can better condition the neural expressive TTS models. We test this hypothesis by incorporating an Information sieve style encoder, previously shown effective, into a vanilla Tacotron. The whole structure of the proposed model is illustrated in Figure \ref{fig:modelstrcuture}.

To the best of the authors’ knowledge, few researchers have investigated the effectiveness of an information sieve in expressive neural TTS models. The closest to this work is CopyCat\cite{dblp:conf/interspeech/CopyCat}, which has a similar structure but accepts a VAE structure’s output as the input, with the VAE potentially working as an information bottleneck. Meanwhile, its style encoder accepts mel-spectrograms, source--speaker embeddings, and phoneme embeddings simultaneously as its input, undermining the effectiveness of a pure information sieve.

\subsection{Definitions and Model Framework}
As general mathematical annotations, upper-case letters such as $X$ refer to random variables or vectors, and lower-case letters to deterministic values or instances of random variables. $X(1 \rightarrow T)$ refers to a random process, with $(1 \rightarrow T)$ denoting a collection of time indices running from $1$ to $T$. We used $E_c(\cdot)$ to denote the text encoder, $E_s(\cdot)$ to denote the style encoder, and $D(\cdot, \cdot)$ to denote the decoder receiving output from the two encoders. The inputs to these modules would be different during training and inference.

\textbf{Training}: We denoted the input to the style encoder as $X$ and that to the text encoder as $Y$, with $X$ and $Y$ being parallel to each other. Then, for each pair of input, we have
\begin{equation}
    C_1 = E_c(X), \ S_1 = E_s(Y), \ \hat{X}_{1\rightarrow 1} = D(C_1, S_1).
    \label{eq1}
\end{equation}
Here, $C_1$, $S_1$, and $\hat{X}_{1\rightarrow 1}$ are random processes. The model learns to reconstruct itself by reconstruction loss, which in turn consists of linear loss and mel loss.
\begin{equation}
    \min\limits_{D(\cdot, \cdot)} = L_{linear\_recon} + L_{mel\_recon}
    \label{eq2}
\end{equation}
The loss is the same with the original GST-Tacotron.

\textbf{Inferring}: During actual synthesis, text $X_1$ is fed into the text encoder, and a random reference audio $Y_2$ is into the style encoder. $X_1$ and $Y_2$ are not parallel to each other. The decoder receives both inputs to synthesize the audio of a certain style.
\begin{equation}
    C_1 = E_c(X_1), \ S_2 = E_s(Y_2), \ \hat{X}_{1\rightarrow 2} = D(C_1, S_2).
    \label{eq3}
\end{equation}
where $C_1$ and $\hat{X}_{1\rightarrow 2}$ are both random processes, and $S_2$ is a reference vector extracted by $E_s(\cdot)$.

\subsection{Sieve Reference Encoder}
\begin{figure}[t]
  \centering
  \includegraphics[width=6cm]{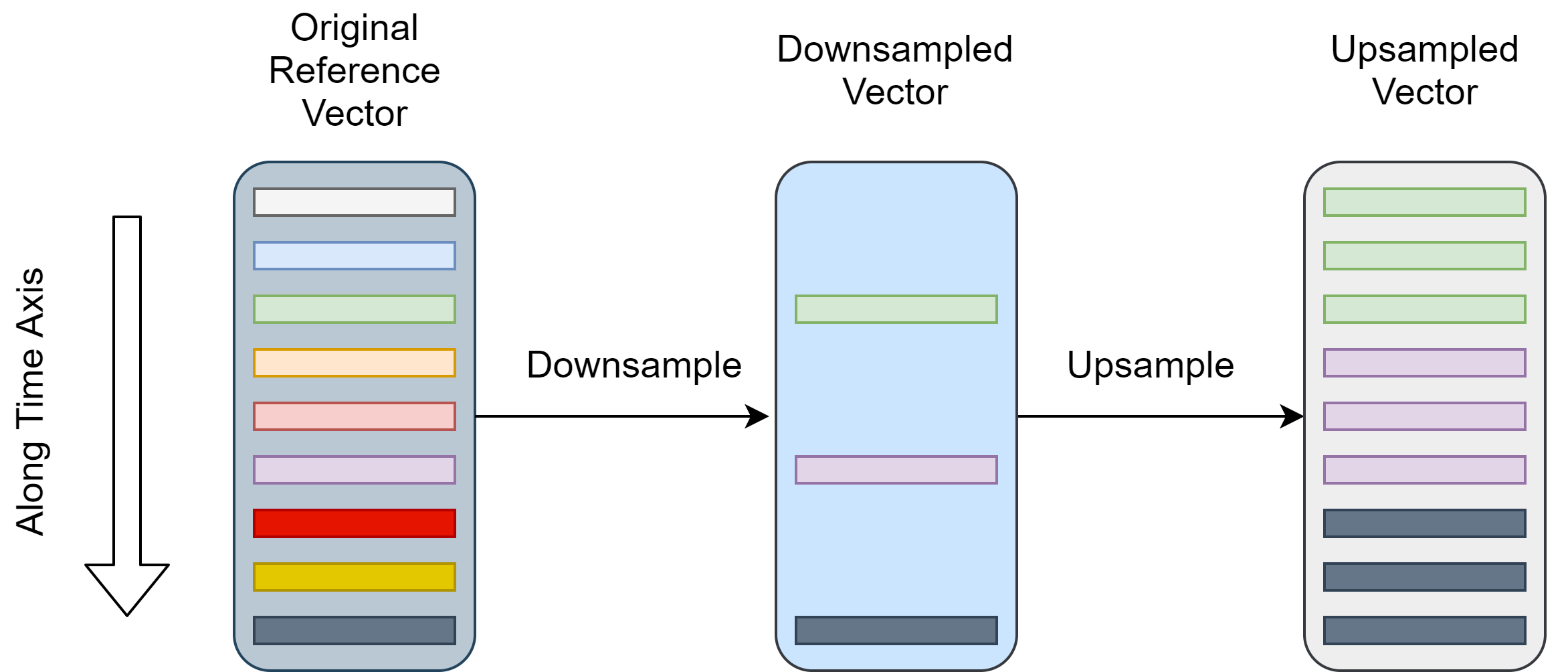}
  \caption{Structure of our sieve layer. For the convenience of demonstration, $\tau$ is set to 3}
  \label{fig:bottleneck}
\end{figure}
We reemphasize our goal to solve content leakage. During the training, the reference encoder learns to extract style from a mel-spectrogram $X$, and the dimension of its output $Z$ controls the volume of information it carries. When dealing with content leakage, we found it problematic to directly reduce the output $Z$ dimension as it worsened the quality of the synthesized audio. While increasing the $Z$ dimension, the leaked content led to apparent errors in the synthesized audio, such as inserted words, missed words, etc. Therefore, we introduce an information sieve, limiting the information flown out from the style encoder to the decoder along the time axis. It compels the decoder to rely on the text encoder to reconstruct the voice content and depend on the sieve style encoder to reconstruct the style.

We downsampled $Z$ along the time axis at a fixed rate $\tau$ to derive $Z^\downarrow \in \mathbb{R}^{\lceil T / \tau \rceil \times H}$. Then, we upsampled $Z^\downarrow$ to length T to reconstruct $\hat{Z} = \mathbb{R}^{T \times H}$. Unlike AutoVC, which downsamples from a bidirectional LSTM unit’s hidden states, this method downsampled from a unidirectional GRU unit’s hidden states based on the assumption that post-order words do not affect pre-ordered words in an auto-regressive model. Hence, we used $\hat{Z}$, which picks out timestamps like $\{\tau - 1, 2\tau - 1, ... T - 1\}$ to retain key information. The structure is depicted in Figure \ref{fig:bottleneck}.

\subsection{Instance Normalization with Conv2D}
It is based on the hypothesis that prosody varies with the channel, while batch normalization will over-smooth the prosody of different audio. Hence, we assume that prosody can be better retained by normalizing it with each channel’s mean and standard deviation but not the batch’s. Therefore, we normalized a set of two-dimensional (2D) convolutional layers by expressing the mean and deviation as follows:
\begin{equation}
    \mu_{ti} = \frac{1}{HW} \sum\limits_{w=1}^W \sum\limits_{h=1}^H  x_{tiwh} \ , \forall w \in W, \forall h \in H
    \label{eq4}
\end{equation}
\begin{equation}
    \sigma_{ti}^2 = \frac{1}{HW} \sum\limits_{l=1}^W \sum\limits_{m=1}^H (x_{tiwh} - \mu_{ti})^2, \ \forall w \in W,\ \forall h \in H
    \label{eq5}
\end{equation}
where $H$ is the length of the time axis input, $W$ is the length of the axis of the mel channels, $w$ is the $w$th step for a kernel along the time axis, and $h$ is the $h$th step along the mel-channel axis. $t$ is the length of the convolution kernel along the time axis, $i$ is the length of the kernel along the mel-channel axis, and $x$ is the convolutional result defined by $t, i, w, h$. Then, we normalized each element by
\begin{equation}
x_{tiwh}' = \frac{{x_{tiwh} - \mu_{ti}}}{\sigma_{ti}}, \ \forall w \in W,\ \forall h \in H
\label{eq6}
\end{equation}

We applied instance normalization to all the six convolutional layers of the style encoder to strengthen its effect.
\begin{figure}[t]
  \centering
  \includegraphics[width=6cm]{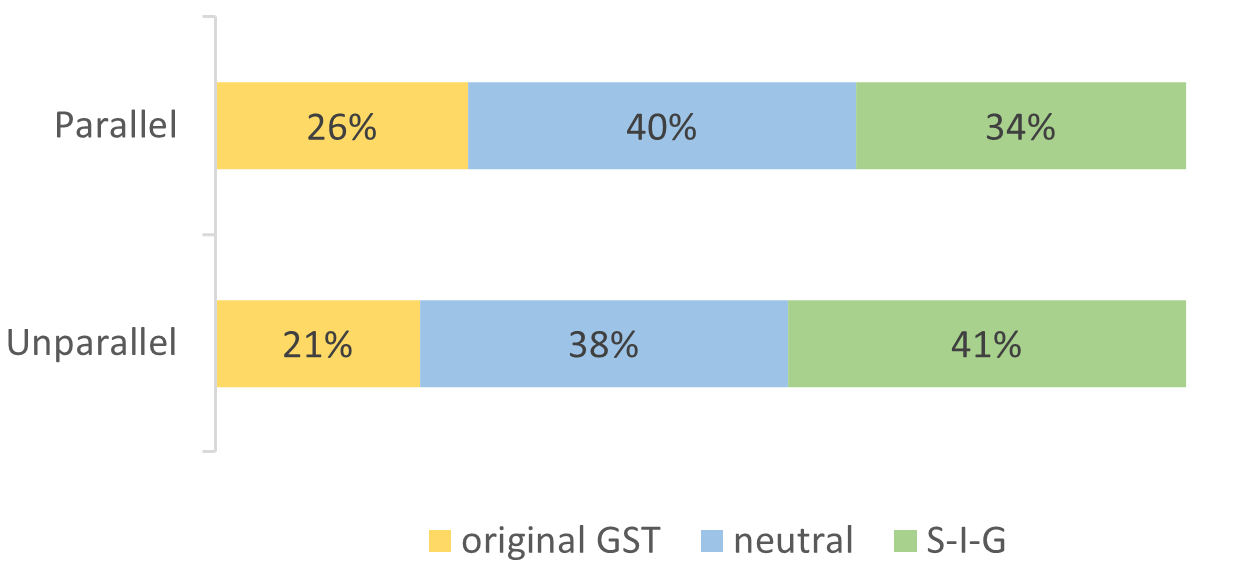}
  \caption{AXY test results for prosody transfer}
  \label{fig:prosodyTransfer}
\end{figure}
\section{Experiments}
\subsection{Dataset and Preprocessing}
The dataset for this study was collected from a public expressive dataset called Blizzard Challenge 2013 (BC2013), which is a professionally recorded, female, single-speaker expressive speech in English. For the sake of accuracy, we only used its segmented part, containing a total of 9,733 valid (text-utterance) pairs.
During the training and inference, we did not incorporate any form of style-related label information.

We divided the 9,733 pairs into 3 parts, 9,550 pairs for training, 50 pairs for validation, and 133 pairs for testing. All the evaluations were conducted on texts from the test set. The mel-spectrograms were computed using 80 mel bins, 50-ms window size, and 12.5-ms hop size. We previously scaled the audio to a range of [-4, 4] in the log-scale. We eliminated all of the silence to enable the audio to sound clear. We transformed raw text input into ArpaBet phonemes with ‘sp’ as pause and ‘sil’ as the end of sentences using an internal front-end model.

As described above, we experimented with the vanilla Tacotron combined with the proposed refined style encoder. The IS-GST follows the same training procedure as the standard Tacotron. We set the learning rate to 0.001, batch size to 16, and $\tau$ to 32. We trained each model in approximately 180,000 steps, which ensures the model converges while preventing overfitting. The conditioning inputs for the training were obtained from the ground-truth audio paired with textual inputs. During inference, a reference can be randomly chosen from the training, validation, and test sets. Users can also manually set the weight for each token. Finally, we used a MelGAN \cite{dblp:conf/nips/melgan} vocoder to generate waveforms from the predicted mel-spectrograms. The vocoder was separately trained on the ground-truth training data from the BC 2013 dataset.
\subsection{Results and Evaluation}
We evaluated the proposed models for prosody transfer using both subjective listening tests and objective metrics. To further test each modification, we conducted an ablation test and compared the results with the ground-truth audio. We list the sources of the samples below using their abbreviations.
\begin{itemize}
\item Ground Truth: Ground-truth audio from the test set.
\item Original GST-Tacotron(GST): GST-Tacotron model \cite{dblp:conf/icml/styletokens}.
\item ASR-guided-Tacotron (ASR-G): shown in \cite{dblp:conf/slt/asrguidedtacotron}.
\item Sieve GST-Tacotron(Sieve GST): GST model’s style encoder upgraded with an information sieve layer.
\item Instance Normalization GST-Tacotron(I-G): Simply replace batch normalization used in GST-Tacotron's style encoders convolution layers with instance normalization.
\item Sieve instance normalization GST-Tacotron (S-I-G): GST model’s style encoder with an information sieve layer; meanwhile, its convolution layers use instance normalization instead of batch normalization.
\end{itemize}

\subsubsection{Subjective Evaluations}
\begin{table}[t]
  \caption{MOS Naturalness scores for the models in 95\% confidence interval}
  \label{tab:MOS-score}
  \centering
  \begin{tabular}{ ll }
   \toprule
   \textbf{Model} & \textbf{MOS score} \\
    \midrule
    Ground Truth         & 4.22 $\pm$ 0.019  \\
    GST                  & 3.54 $\pm$ 0.024  \\
    ASR-G                & 3.93 $\pm$ 0.041  \\
    Sieve GST           & 3.82 $\pm$ 0.031  \\
    I-G                 & 3.68 $\pm$ 0.028  \\
    S-I-G               & 3.91 $\pm$ 0.029   \\
    \bottomrule
  \end{tabular}
\end{table}

\begin{figure}[t]
  \centering
  \includegraphics[width=6.5cm]{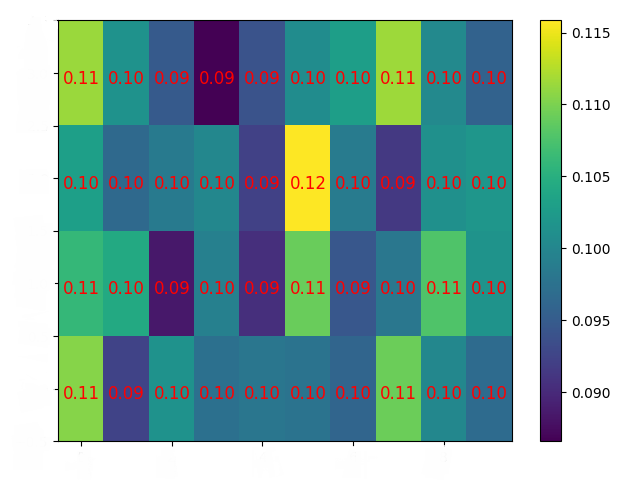}
  \caption{This figure demonstrates a common phenomenon while we were trying to reproduce the original GST-Tacotron, that is the average weight per token. This is even harder for us to distinguish each token's unique effect since we incorporate multi-head attention to refine results.}
  \label{fig:original}
\end{figure}
We begin with a mean opinion score (MOS) evaluation of speech naturalness. We used texts from the test set and generated speech using the aforementioned models compared with the ground truth test audio. For each MOS test, a total of 20 English-speaking listeners were invited. Each listener was presented with 10 samples from each source and then asked to rate the naturalness of the audio on a scale of 1--5 in intervals of 0.5. The results are listed in Table \ref{tab:MOS-score}. The information sieve model was rated more natural than the original GST model. The I-G model did not benefit significantly from instance normalization as it could not effectively eliminate mispronunciations. When combined, the positive effects of instance normalization and information sieve were coamplified as to get close to the MOS of the SOTA ASR-G. \footnote{According to the survey, listeners gave the new models a higher rating on the account of it having fewer incorrectly repeated or missed words compared with the audio synthesized by the original GST model.} To view the demonstration of the same, visit \url{https://hsudongdai.github.io/interspeech_2021_demo/}.

Next, we evaluated the prosody transfer from the reference audio to the synthesized speech for an arbitrary sentence. For convenience, we merely considered the synthesized audio from the S-I-G model and the original GST model to conduct a side-by-side test. We conducted parallel prosody transfer, where texts and audio are one-to-one matched, and unparallel prosody transfer, where texts and audio are randomly paired. We constructed triplets (A, X, Y), where A is the reference audio, and X and Y are utterances synthesized using the same text but by S-I-G and the original GST model, respectively. For each test, 10 English-speaking listeners were invited and presented with ten triplets. They were asked to choose the one more similar to the prosody of the reference audio. Figure \ref{fig:prosodyTransfer} illustrates the results of the listening tests. The S-I-G model transfers prosody better than the original model, according to the higher percentage of votes obtained. Because this model is a CPT model without fine-grained features, it is an acceptable result. Moreover, it still surpasses the original GST model.

\subsubsection{Objective Evaluations}
The objectives of this model were to mitigate content leakage in the original GST Tacotron and make the tokens in the GST layer more specific and orthogonal to one another for better style control. Hence, we conducted two objective evaluations.
\begin{enumerate}
\item Compute the WER and word information lost (WIL) \cite{dblp:conf/interspeech/wer&wil} for all the models trained.
\item Use audio of different styles to demonstrate the orthogonality of the style tokens.
\end{enumerate}
\begin{figure}[t]
  \centering
  \includegraphics[width=6.5cm]{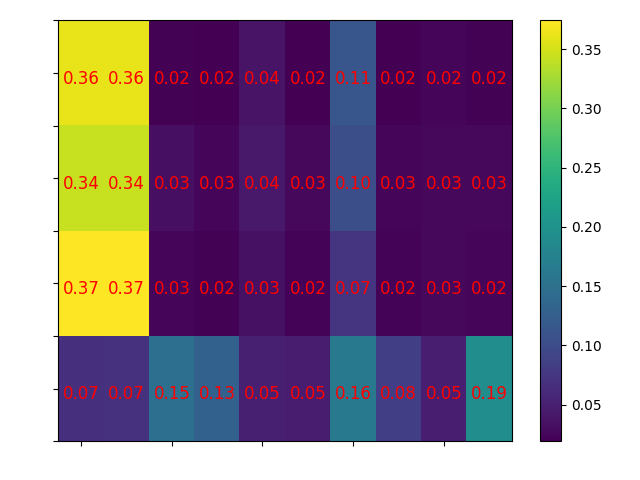}
  \caption{This figure demonstrates tokens' weights generated by our sieve style encoder. Obviously, we can observe 1st and 2nd token win more weight. Similar results are omitted due to the limit of pages.}
  \label{fig:refined}
\end{figure}
For the first experiment, we selected 50 sentences from the test set, the ground-truth audio of which were 3--7 s long, to compute the WER. A reference neutral prosody was used uniformly for all models and the 50 sentences. The results are listed in Table \ref{tab:WER-score}. Clearly, the WERs of S-G and S-I-G models were merely 30\% of that of the original GST model, whereas that of the I-G model was approximately the same with the original GST models, proving that instance normalization does not contribute to mitigating content leakage. This result proves the effectiveness of the information sieve. To further prove the accuracy of this model, we computed the WIL, which reflects similar results.
\begin{table}[tp]
  \caption{WER(Word Error Rate) and WIL (Word Information Lost) for each model. We use a PyPi package called jiwer to calculate both indices. Though widely used as a metric for ASR, it also reflects the stability of our TTS model.}
  \label{tab:WER-score}
  \centering
  \begin{tabular}{ cccccc }
    \toprule
   Model & GST & ASR-G & Sieve GST &  I-G & S-I-G \\
    \midrule
   WER & 0.2942 & 0.2453 & 0.1133 & 0.2587 & 0.1163 \\
   WIL & 0.4218 & 0.3734 & 0.1876 & 0.2413 & 0.1995 \\
    \bottomrule
  \end{tabular}
\end{table}
In the second experiment, we selected several pieces of audio of the \textbf{Strong} style to demonstrate the weight of tokens per reference audio. Compared with the almost equal weight distribution of the original GST-Tacotron in Figure \ref{fig:original}, the information sieve model occasionally demonstrates a stronger allocation preference for certain tokens, assigning much more weight than other inactivated tokens when accepting different pieces of reference audio, which helps distinguish each token shown in Figure \ref{fig:refined}.

\section{Conclusions}
We demonstrated a simple yet effective method that mitigates content leakage while still synthesizing highly natural, expressive speech and conducted style transfer. By adding a simple information sieve layer on a style encoder to condition a neural TTS model, we could transfer prosody from the reference audio, or a manual combination of several tokens, to any synthesized speech. The sieve successfully tackled content leakage while preserving key style information. 
Furthermore, we shall focus on searching for better parameters to control the sieve to accurately retain useful information while rejecting the rest for any given mission and building a mathematical theory to better explain the sieve structure.

\section{Acknowledgements}
This work was supported by the National Key R\&D Program of China under Grant 2018YFB1305200, the Tianjin Municipal Science and Technology Project under Grant 18ZXZNGX00330 and the National Natural
Science Foundation of China under Grant 61771333.

\bibliographystyle{IEEEtran}

\bibliography{mybib}

\begin{thebibliography}{10}
\providecommand{\url}[1]{#1}
\csname url@samestyle\endcsname
\providecommand{\newblock}{\relax}
\providecommand{\bibinfo}[2]{#2}
\providecommand{\BIBentrySTDinterwordspacing}{\spaceskip=0pt\relax}
\providecommand{\BIBentryALTinterwordstretchfactor}{4}
\providecommand{\BIBentryALTinterwordspacing}{\spaceskip=\fontdimen2\font plus
\BIBentryALTinterwordstretchfactor\fontdimen3\font minus
  \fontdimen4\font\relax}
\providecommand{\BIBforeignlanguage}[2]{{%
\expandafter\ifx\csname l@#1\endcsname\relax
\typeout{** WARNING: IEEEtran.bst: No hyphenation pattern has been}%
\typeout{** loaded for the language `#1'. Using the pattern for}%
\typeout{** the default language instead.}%
\else
\language=\csname l@#1\endcsname
\fi
#2}}
\providecommand{\BIBdecl}{\relax}
\BIBdecl

\bibitem{dblp:conf/interspeech/WangTacotron}
Y.~Wang, R.~J. Skerry{-}Ryan, D.~Stanton, Y.~Wu, R.~J. Weiss, N.~Jaitly,
  Z.~Yang, Y.~Xiao, Z.~Chen, S.~Bengio, Q.~V. Le, Y.~Agiomyrgiannakis,
  R.~Clark, and R.~A. Saurous, ``Tacotron: Towards end-to-end speech
  synthesis,'' in \emph{Interspeech 2017, 18th Annual Conference of the
  International Speech Communication Association}.\hskip 1em plus 0.5em minus
  0.4em\relax {ISCA}, 2017, pp. 4006--4010.

\bibitem{dblp:conf/icassp/ShenPWSJYCZWRSA18}
J.~Shen, R.~Pang, R.~J. Weiss, M.~Schuster, N.~Jaitly, Z.~Yang, Z.~Chen,
  Y.~Zhang, Y.~Wang, R.~Ryan, R.~A. Saurous, Y.~Agiomyrgiannakis, and Y.~Wu,
  ``Natural {TTS} synthesis by conditioning wavenet on {MEL} spectrogram
  predictions,'' in \emph{2018 {IEEE} International Conference on Acoustics,
  Speech and Signal Processing, {ICASSP} 2018}.\hskip 1em plus 0.5em minus
  0.4em\relax {IEEE}, 2018, pp. 4779--4783.

\bibitem{dblp:conf/nips/RenRTQZZL19}
Y.~Ren, Y.~Ruan, X.~Tan, T.~Qin, S.~Zhao, Z.~Zhao, and T.~Liu, ``Fastspeech:
  Fast, robust and controllable text to speech,'' in \emph{Advances in Neural
  Information Processing Systems 32: Annual Conference on Neural Information
  Processing Systems 2019, NeurIPS 2019,}, 2019, pp. 3165--3174.

\bibitem{dblp:conf/iclr/deepvoice3}
W.~Ping, K.~Peng, A.~Gibiansky, S.~{\"{O}}. Arik, A.~Kannan, S.~Narang,
  J.~Raiman, and J.~Miller, ``Deep voice 3: Scaling text-to-speech with
  convolutional sequence learning,'' in \emph{6th International Conference on
  Learning Representations, {ICLR} 2018, Vancouver, BC, Canada, April 30 - May
  3, 2018, Conference Track Proceedings}.\hskip 1em plus 0.5em minus
  0.4em\relax OpenReview.net, 2018.

\bibitem{gong2021improving}
C.~Gong, L.~Wang, Z.~Ling, S.~Guo, J.~Zhang, and J.~Dang, ``Improving
  naturalness and controllability of sequence-to-sequence speech synthesis by
  learning local prosody representations,'' \emph{ICASSP 2021-2021 IEEE
  International Conference on Acoustics, Speech and Signal Processing
  (ICASSP)}, pp. 5724--5728, 2021.

\bibitem{dblp:conf/icml/towardse2e}
R.~J. Skerry{-}Ryan, E.~Battenberg, Y.~Xiao, Y.~Wang, D.~Stanton, J.~Shor,
  R.~J. Weiss, R.~Clark, and R.~A. Saurous, ``Towards end-to-end prosody
  transfer for expressive speech synthesis with tacotron,'' in
  \emph{Proceedings of the 35th International Conference on Machine Learning,
  {ICML} 2018, Stockholmsm{\"{a}}ssan, Stockholm, Sweden, July 10-15, 2018},
  vol.~80.\hskip 1em plus 0.5em minus 0.4em\relax {PMLR}, 2018, pp. 4700--4709.

\bibitem{dblp:conf/icml/styletokens}
Y.~Wang, D.~Stanton, Y.~Zhang, R.~J. Skerry{-}Ryan, E.~Battenberg, J.~Shor,
  Y.~Xiao, Y.~Jia, F.~Ren, and R.~A. Saurous, ``Style tokens: Unsupervised
  style modeling, control and transfer in end-to-end speech synthesis,'' in
  \emph{Proceedings of the 35th International Conference on Machine Learning,
  {ICML} 2018, Stockholmsm{\"{a}}ssan, Stockholm, Sweden, July 10-15, 2018},
  vol.~80.\hskip 1em plus 0.5em minus 0.4em\relax {PMLR}, 2018, pp. 5167--5176.

\bibitem{dblp:conf/icassp/ZhangPHL19}
Y.~Zhang, S.~Pan, L.~He, and Z.~Ling, ``Learning latent representations for
  style control and transfer in end-to-end speech synthesis,'' in \emph{{IEEE}
  International Conference on Acoustics, Speech and Signal Processing, {ICASSP}
  2019, Brighton, United Kingdom, May 12-17, 2019}.\hskip 1em plus 0.5em minus
  0.4em\relax {IEEE}, 2019, pp. 6945--6949.

\bibitem{dblp:conf/interspeech/AkuzawaIM18}
K.~Akuzawa, Y.~Iwasawa, and Y.~Matsuo, ``Expressive speech synthesis via
  modeling expressions with variational autoencoder,'' in \emph{Interspeech
  2018, 19th Annual Conference of the International Speech Communication
  Association, Hyderabad, India, 2-6 September 2018}.\hskip 1em plus 0.5em
  minus 0.4em\relax {ISCA}, 2018, pp. 3067--3071.

\bibitem{dblp:conf/iclr/HsuHierarchical}
W.~Hsu, Y.~Zhang, R.~J. Weiss, H.~Zen, Y.~Wu, Y.~Wang, Y.~Cao, Y.~Jia, Z.~Chen,
  J.~Shen, P.~Nguyen, and R.~Pang, ``Hierarchical generative modeling for
  controllable speech synthesis,'' in \emph{7th International Conference on
  Learning Representations, {ICLR} 2019, New Orleans, LA, USA, May 6-9,
  2019}.\hskip 1em plus 0.5em minus 0.4em\relax OpenReview.net, 2019.

\bibitem{dblp:conf/interspeech/KlimkovHierarchical}
V.~Klimkov, S.~Ronanki, J.~Rohnke, and T.~Drugman, ``Fine-grained robust
  prosody transfer for single-speaker neural text-to-speech,'' in
  \emph{Interspeech 2019, 20th Annual Conference of the International Speech
  Communication Association, Graz, Austria, 15-19 September 2019}.\hskip 1em
  plus 0.5em minus 0.4em\relax {ISCA}, 2019, pp. 4440--4444.

\bibitem{dblp:conf/icassp/SunHierarchical}
G.~Sun, Y.~Zhang, R.~J. Weiss, Y.~Cao, H.~Zen, and Y.~Wu, ``Fully-hierarchical
  fine-grained prosody modeling for interpretable speech synthesis,'' in
  \emph{2020 {IEEE} International Conference on Acoustics, Speech and Signal
  Processing, {ICASSP} 2020, Barcelona, Spain, May 4-8, 2020}.\hskip 1em plus
  0.5em minus 0.4em\relax {IEEE}, 2020, pp. 6264--6268.

\bibitem{dblp:conf/interspeech/HonoHierarchical}
Y.~Hono, K.~Tsuboi, K.~Sawada, K.~Hashimoto, K.~Oura, Y.~Nankaku, and
  K.~Tokuda, ``Hierarchical multi-grained generative model for expressive
  speech synthesis,'' in \emph{Interspeech 2020, 21st Annual Conference of the
  International Speech Communication Association, Virtual Event, Shanghai,
  China, 25-29 October 2020}.\hskip 1em plus 0.5em minus 0.4em\relax {ISCA},
  2020, pp. 3441--3445.

\bibitem{dblp:conf/icassp/HuMutualInformation}
T.~Hu, A.~Shrivastava, O.~Tuzel, and C.~Dhir, ``Unsupervised style and content
  separation by minimizing mutual information for speech synthesis,'' in
  \emph{2020 {IEEE} International Conference on Acoustics, Speech and Signal
  Processing, {ICASSP} 2020, Barcelona, Spain, May 4-8, 2020}.\hskip 1em plus
  0.5em minus 0.4em\relax {IEEE}, 2020, pp. 3267--3271.

\bibitem{dblp:conf/slt/asrguidedtacotron}
D.~Liu, C.~Yang, S.~Wu, and H.~Lee, ``Improving unsupervised style transfer in
  end-to-end speech synthesis with end-to-end speech recognition,'' in
  \emph{2018 {IEEE} Spoken Language Technology Workshop, {SLT} 2018, Athens,
  Greece, December 18-21, 2018}.\hskip 1em plus 0.5em minus 0.4em\relax {IEEE},
  2018, pp. 640--647.

\bibitem{dblp:journals/corr/gururani}
S.~Gururani, K.~Gupta, D.~Shah, Z.~Shakeri, and J.~Pinto, ``Prosody transfer in
  neural text to speech using global pitch and loudness features,''
  \emph{CoRR}, vol. abs/1911.09645, 2019.

\bibitem{dblp:conf/ccs/SongR20}
C.~Song and A.~Raghunathan, ``Information leakage in embedding models,'' in
  \emph{{CCS} '20: 2020 {ACM} {SIGSAC} Conference on Computer and
  Communications Security, Virtual Event, USA, November 9-13, 2020}.\hskip 1em
  plus 0.5em minus 0.4em\relax {ACM}, 2020, pp. 377--390.

\bibitem{dblp:journals/corr/cvcontentleakage}
K.~Ridgeway and M.~C. Mozer, ``Open-ended content-style recombination via
  leakage filtering,'' \emph{CoRR}, vol. abs/1810.00110, 2018.

\bibitem{dblp:conf/cvpr/RoyContentLeakage}
P.~C. Roy and V.~N. Boddeti, ``Mitigating information leakage in image
  representations: {A} maximum entropy approach,'' in \emph{{IEEE} Conference
  on Computer Vision and Pattern Recognition, {CVPR} 2019, Long Beach, CA, USA,
  June 16-20, 2019}.\hskip 1em plus 0.5em minus 0.4em\relax Computer Vision
  Foundation / {IEEE}, 2019, pp. 2586--2594.

\bibitem{dblp:conf/icml/QianAutoVC}
K.~Qian, Y.~Zhang, S.~Chang, X.~Yang, and M.~Hasegawa{-}Johnson, ``Autovc:
  Zero-shot voice style transfer with only autoencoder loss,'' in
  \emph{Proceedings of the 36th International Conference on Machine Learning,
  {ICML} 2019, 9-15 June 2019, Long Beach, California, {USA}}, vol.~97.\hskip
  1em plus 0.5em minus 0.4em\relax {PMLR}, 2019, pp. 5210--5219.

\bibitem{dblp:conf/interspeech/CopyCat}
S.~Karlapati, A.~Moinet, A.~Joly, V.~Klimkov, D.~S{\'{a}}ez{-}Trigueros, and
  T.~Drugman, ``Copycat: Many-to-many fine-grained prosody transfer for neural
  text-to-speech,'' in \emph{Interspeech 2020, 21st Annual Conference of the
  International Speech Communication Association, Virtual Event, Shanghai,
  China, 25-29 October 2020}.\hskip 1em plus 0.5em minus 0.4em\relax {ISCA},
  2020, pp. 4387--4391.

\bibitem{dblp:conf/nips/melgan}
K.~Kumar, R.~Kumar, T.~de~Boissiere, L.~Gestin, W.~Z. Teoh, J.~Sotelo,
  A.~de~Br{\'{e}}bisson, Y.~Bengio, and A.~C. Courville, ``Melgan: Generative
  adversarial networks for conditional waveform synthesis,'' in \emph{Advances
  in Neural Information Processing Systems 32: Annual Conference on Neural
  Information Processing Systems 2019, NeurIPS 2019, December 8-14, 2019,
  Vancouver, BC, Canada}, 2019, pp. 14\,881--14\,892.

\bibitem{dblp:conf/interspeech/wer&wil}
A.~C. Morris, V.~Maier, and P.~D. Green, ``From {WER} and {RIL} to {MER} and
  {WIL:} improved evaluation measures for connected speech recognition,'' in
  \emph{{INTERSPEECH} 2004 - ICSLP, 8th International Conference on Spoken
  Language Processing, Jeju Island, Korea, October 4-8, 2004}.\hskip 1em plus
  0.5em minus 0.4em\relax {ISCA}, 2004.

\end{thebibliography}

\end{document}